\documentclass[aps,prb,twocolumn,superscriptaddress,showpacs]{revtex4}

\usepackage{graphicx}% Include figure files
\usepackage{dcolumn}% Align table columns on decimal point
\usepackage{bm}% bold math

\begin{document}

\title{Large magnetothermal conductivity of HoMnO$_3$ single crystals
and its relation to the magnetic-field induced transitions of
magnetic structure}

\author{X. M. Wang}
\affiliation{Hefei National Laboratory for Physical Sciences at
Microscale, University of Science and Technology of China, Hefei,
Anhui 230026, People's Republic of China}

\author{C. Fan}
\affiliation{Hefei National Laboratory for Physical Sciences at
Microscale, University of Science and Technology of China, Hefei,
Anhui 230026, People's Republic of China}

\author{Z. Y. Zhao}
\affiliation{Hefei National Laboratory for Physical Sciences at
Microscale, University of Science and Technology of China, Hefei,
Anhui 230026, People's Republic of China}

\author{W. Tao}
\affiliation{Hefei National Laboratory for Physical Sciences at
Microscale, University of Science and Technology of China, Hefei,
Anhui 230026, People's Republic of China}

\author{X. G. Liu}
\affiliation{Hefei National Laboratory for Physical Sciences at
Microscale, University of Science and Technology of China, Hefei,
Anhui 230026, People's Republic of China}

\author{W. P. Ke}
\affiliation{Hefei National Laboratory for Physical Sciences at
Microscale, University of Science and Technology of China, Hefei,
Anhui 230026, People's Republic of China}

\author{X. Zhao}
\affiliation{School of Physical Sciences, University of Science
and Technology of China, Hefei, Anhui 230026, People's Republic of
China}

\author{X. F. Sun}
\email{xfsun@ustc.edu.cn}

\affiliation{Hefei National Laboratory for Physical Sciences at
Microscale, University of Science and Technology of China, Hefei,
Anhui 230026, People's Republic of China}

\date{\today}

\begin{abstract}

We study the low-temperature heat transport of HoMnO$_3$ single
crystals to probe the magnetic structures and their transitions
induced by magnetic field. It is found that the low-$T$ thermal
conductivity ($\kappa$) shows very strong magnetic-field
dependence, with the strongest suppression of nearly 90\% and the
biggest increase of 20 times of $\kappa$ compared to its
zero-field value. In particular, some ``dip"-like features show up
in $\kappa(H)$ isotherms for field along both the $ab$ plane and
the $c$ axis. These behaviors are found to shed new light on the
complex $H$-$T$ phase diagram and the field-induced
re-orientations of Mn$^{3+}$ and Ho$^{3+}$ spin structures. The
results also demonstrate a significant spin-phonon coupling in
this multiferroic compound.

\end{abstract}

%66.70.-f Nonelectronic thermal conduction and heat-pulse propagation in solids
%75.47.-m Magnetotransport phenomena; materials for magnetotransport
%75.50.-y Studies of specific magnetic materials
%75.85.+t Magnetoelectric effects, multiferroics

\maketitle

\section{Introduction}

Magnetic-field-induced transition of magnetism is an outstanding
phenomenon in the strongly-correlated electron systems and is
associated with many physical interests, such as the
unconventional superconductivity,\cite{Lake, Kang, Chang} the
non-Fermi-liquid behaviors,\cite{Custers} and the
multiferroicity,\cite{Kimura} etc. It is known that
multiferroicity is a result of strong coupling between magnetic
and electric degrees of freedom in insulators and has received a
lot of research interests because of its application usage. In
this family of materials, for example the rare-earth manganites
RMnO$_3$ (R = rare-earth elements),\cite{Kimura, Aken, Ueland,
Lottermoser} it is commonly found that the magnetic-field-induced
transitions of magnetic structures are accompanied by the drastic
changes of dielectric properties. Understanding the microscopic
magnetic ordering across these transition boundaries is therefore
very helpful for revealing the mechanism of magnetoelectric
coupling, a topic of much current interest.

We select the hexagonal manganite HoMnO$_3$ as a representative
object for studying the magnetic-field-induced transitions and
their impacts on physical properties. This compound is
ferroelectric below $T_c$ = 875 K and the moments of Mn$^{3+}$ and
Ho$^{3+}$ ions display antiferromagnetic (AF) orderings at
$T_{N,Mn}$ = 75 K and $T_{N,Ho}$ = 4.6 K, respectively. The
peculiarity of HoMnO$_3$ is that it shows several transitions of
the magnetic structure upon lowering temperature or increasing
magnetic field, which results in a very rich $H$-$T$ phase diagram
at low temperatures.\cite{Lottermoser, Vajk, Lorenz, Yen, Lemyre}
Furthermore, these low-$T$ field-induced transitions are confirmed
to cause drastic changes of dielectric constant.\cite{Yen}
However, the low-$T$ magnetic structures and the mechanisms of
field-induced transitions are still not fully understood, in spite
of a lot of experimental investigations using neutron scattering,
optical and microwave techniques, etc.\cite{Lonkai, Fiebig, Brown,
Nandi, Hur} This is due to the complexity of Mn$^{3+}$ and
Ho$^{3+}$ magnetism in this compound. The Mn$^{3+}$ ions form
triangular planar sublattices and the AF exchange coupling among
Mn$^{3+}$ moments is geometrically frustrated. As a result, below
$T_{N,Mn}$, the Mn$^{3+}$ moments are ordered in a configuration
that the neighboring moments are 120$^{\circ}$ rotated. However,
the homometric configurations of the Mn$^{3+}$ moments in the
triangular lattice are difficult to distinguish from each
other.\cite{Lottermoser, Vajk, Lorenz, Yen, Lemyre, Lonkai,
Fiebig, Brown, Nandi, Hur} It was assumed that the Ho$^{3+}$
moments are orientated along the $c$ axis with an Ising-like
anisotropy, and their spin structures are not simple because the
Ho$^{3+}$ ions present in two different crystallographic sites.
The formed two Ho$^{3+}$ sublattices can order separately in
either parallel or anti-parallel orientations.\cite{Lottermoser,
Brown, Nandi, Hur} For these complexities, even the zero-field
spin structures at low temperatures have not been completely
clarified, let alone those in the field-induced phases.

Heat transport behavior has recently received a lot of interests
in studying strongly-correlated electron systems and
low-dimensional spin systems.\cite{Hussey, Hess, Sologubenko} It
has been found that the low-$T$ heat transport can probe not only
the transport properties of many kinds of elementary excitations,
such as phonons, magnons, spinons and quasiparitcles in
superconductors, but also the coupling between crystal lattice and
spins.\cite{Sun_PLCO, Sun_GBCO, Sharma} In addition, the heat
transport behavior in magnetic field is an effective way to study
the magnetic-field-induced quantum phase
transitions,\cite{Sun_LSCO, Sun_YBCO, Ando_BSLCO, Paglione,
Sologubenko2} for example, the field-induced AF ordering (which
can be described as the magnon Bose-Einstein condensation) in
spin-gapped quantum magnets.\cite{Sun_DTN} In this paper, we show
that the low-$T$ heat transport of HoMnO$_3$ single crystals is
very useful for probing the magnetic structures and the
field-induced phase transitions. It is found that the spin-phonon
coupling is extremely strong in this material and results in very
strong field dependence of thermal conductivity ($\kappa$). In
particular, the thermal conductivity is significantly suppressed
at several critical fields, which are related to the transitions
of magnetic structure. The present data provide new suggestions on
the magnetic structures of the low-$T$ phases.

\section{Experiments}

High-quality HoMnO$_3$ single crystals are grown by using a
floating-zone technique. The crystals are carefully checked by
using the X-ray Laue photograph and cut precisely along the
crystallographic axes, with parallelepiped shape and typical size
of $2.5 \times 0.6 \times 0.15 mm^3$. The thermal conductivities
are measured along both the $ab$ plane ($\kappa_{ab}$) and the $c$
axis ($\kappa_c$) by using a conventional steady-state technique
and two different processes: (i) using a ``one heater, two
thermometers" technique in a $^3$He refrigerator and a 14 T magnet
at temperature regime of 0.3--8 K; (ii) using a Chromel-Constantan
thermocouple in a $^4$He cryostat for zero-field data above 4
K.\cite{Sun_DTN}

\section{Results and Discussion}

\begin{figure}
\includegraphics[clip,width=6.5cm]{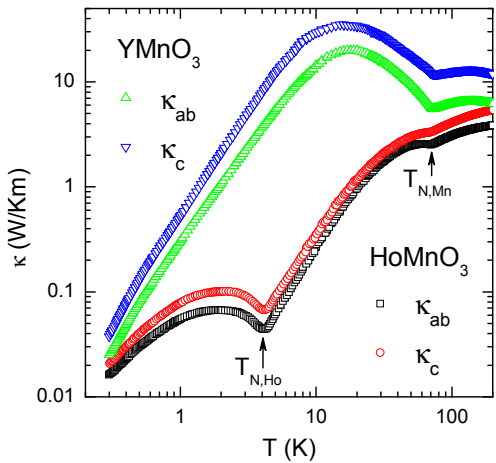}
\caption{(color online) Temperature dependences of the $ab$-plane
and the $c$-axis thermal conductivities of HoMnO$_3$ single
crystals. The heat transport data of YMnO$_3$ single crystals
grown by the floating-zone method are also shown for comparison.}
\end{figure}

Figure 1 shows the temperature dependences of $\kappa_{ab}$ and
$\kappa_{c}$ of HoMnO$_3$ single crystals in zero field, together
with those of YMnO$_3$ for comparison. These data essentially
reproduce the results of a former report.\cite{Sharma} The heat
transport behavior of YMnO$_3$ is a typical one in insulating
crystals,\cite{Berman} except for a ``dip"-like feature at $\sim$
70 K. The large phonon peaks at 15 K indicate a high quality of
the crystals. The ``dip" at 70 K is clearly related to the AF
ordering of Mn$^{3+}$ moments, where the spin fluctuations scatter
phonons strongly. Similar feature at 70 K also appears in
HoMnO$_3$. However, the heat transport of HoMnO$_3$ is very
different from that of YMnO$_3$ in two aspects. First, the low-$T$
phonon conductivity is much weaker in HoMnO$_3$, with the phonon
peak completely being wiped out, suggesting very strong phonon
scattering.\cite{Sharma} This is in good agreement with the
evidences of spin-phonon coupling from other
measurements.\cite{Cruz, Fabreges, Hur} Second, another
``dip"-like feature shows up at $\sim$ 4 K, which corresponds to
the AF ordering temperature of Ho$^{3+}$ moments. It is important
to notice that although Ho$^{3+}$ moments are ordered below 4.6
K,\cite{Lottermoser} the strong phonon scattering seems to appear
at temperature as high as 70 K. Such broad temperature regime for
spin-phonon scattering is not likely to be caused only by the
Ho$^{3+}$ moments. It is already known that the Mn$^{3+}$
sublattice has several spin re-orientation transitions upon
lowering temperature.\cite{Lottermoser, Vajk, Lorenz, Yen, Lemyre}
Obviously, these successive transitions strengthen the spin
fluctuations in Mn$^{3+}$ sublattice and contribute to the phonon
scattering. In contrast, the re-orientations of Mn$^{3+}$ spin
lattice do not occur in YMnO$_3$.

\begin{figure}
\includegraphics[clip,width=8.5cm]{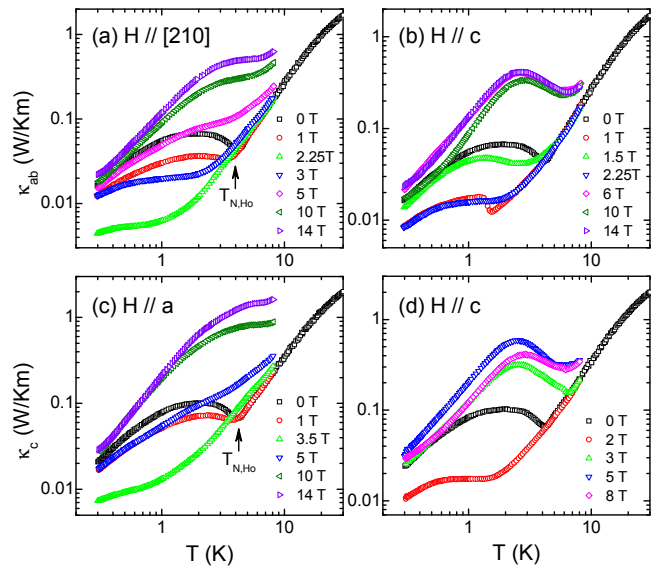}
\caption{(color online) Temperature dependences of thermal
conductivities of HoMnO$_3$ single crystals in both the zero field
and several different magnetic fields up to 14 T. The heat current
are applied along the $ab$ plane or the $c$ axis, while the
magnetic field applied either in the $ab$ plane (along the
$a$-axis or [210] directions) or along the $c$ axis.}
\end{figure}

The strong spin-phonon coupling suggests that the magnetic-field
dependence of heat transport would be very useful for studying the
low-$T$ magnetic structures and their transitions. Figure 2 shows
the temperature dependences of thermal conductivity in different
magnetic fields up to 14 T, which clearly show an extremely large
magnetothermal effect. By passing, it is worthy of pointing out
that the magnetic field produces so strong torque, especially for
$H \parallel c$, that the samples can be completely destroyed.
This makes the measurement very difficult and it is almost
impossible to get high-field data in some field directions. It can
be seen in Fig. 2 that a low magnetic field strongly suppresses
the phonon heat transport and makes the zero-field dip of
$\kappa(T)$ at $T_{N,Ho}$ to be shallower and broader and to shift
to lower temperatures. In high magnetic field, the heat transport
is significantly enhanced, particularly at temperature regime of
2--6 K, in which the spin-phonon scattering seems to be the
strongest in zero field. This clearly demonstrates that the high
magnetic field weakens the phonon scattering by spin fluctuations.
However, one should note that although the field dependence
becomes rather weak in high fields and below 1 K, the temperature
dependence of $\kappa$ in high fields is still much weaker than
the $T^3$ law, suggesting the remaining of some microscopic
scattering on phonons.

\begin{figure}
\includegraphics[clip,width=8.5cm]{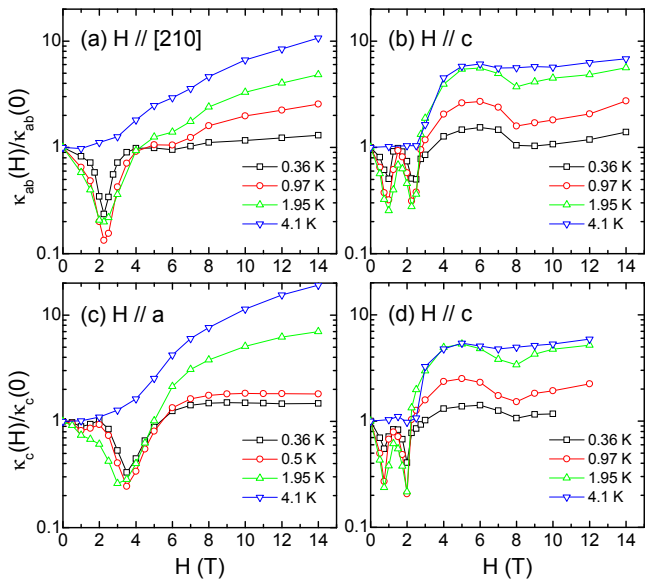}
\caption{(color online) Magnetic-field dependences of thermal
conductivities of HoMnO$_3$ single crystals at low temperatures.}
\end{figure}

The detailed magnetic-field dependences of thermal conductivity
$\kappa(H)$ are shown in Fig. 3. The extremely strong coupling
between phonons and magnetic excitations leads to so strong field
dependence of $\kappa$ that the magnitude of thermal conductivity
can be suppressed down to $\sim$ 10\% at some particular low
fields and be enhanced up to 20 times at 14 T. To our knowledge,
such a large magnetothermal effect has not been observed in other
materials including some rare-earth manganites.\cite{Berggold} At
4.1 K, where the zero-field $\kappa(T)$ data indicate the
strongest spin-phonon scattering, the external magnetic field
suppresses the magnetic excitations and reduces the phonon
scattering. It is therefore easy to understand that both
$\kappa_{ab}$ and $\kappa_c$ show significant enhancement with
increasing field. Apparently, the thermal conductivity can be even
larger before it gets saturated in very high magnetic field above
14 T ($\parallel ab$) and becomes at least comparable to that of
YMnO$_3$, which is free from the spin-phonon scattering.

It is notable that the $\kappa(H)$ isotherms below 4 K are quite
different from the higher-$T$ results. An interesting phenomenon
is that there is one ``dip" in $\kappa(H)$ curves for magnetic
field in the $ab$ plane, while there are two ``dips" in
$\kappa(H)$ for the magnetic field along the $c$ axis. The
characteristic of these ``dips" is that the fields for the minimum
$\kappa$ are nearly independent of the temperature, which
indicates an origin from the spin-flop transitions or spin
re-orientations.\cite{Spin_flop} In an AF ordered state, the
Zeeman energy causes the magnon excitations to become gapless at
the spin-flop field, but the gap opens again at higher
field;\cite{Spin_flop, Sun_PLCO, Jin} consequently, the magnon
scattering of phonons is the strongest at the spin-flop field,
where the magnons are the most populated, and causes a
``dip''-like feature in the $\kappa(H)$ curve.

As we know, the low-$T$ field-induced transitions for $H \parallel
c$ have been intensively studied by the dielectric-constant,
microwave and neutron measurements.\cite{Lorenz, Yen, Lemyre,
Brown} The ``dip" fields of $\kappa(H)$ at $\sim$ 1 and 2 T are in
rather good correspondence with the two successive transitions
among the zero-field phase and some field-induced phases (named as
``LT1" and ``LT2" phases in Refs. [\onlinecite{Lorenz, Yen}]). It
should be pointed out that the low-$T$ magnetic structure has
actually not been completely understood. Between the two
possibilities of the $P6_3cm$ and $P6_3c'm'$ space groups being
the proposed zero-field Mn$^{3+}$ magnetic
structure,\cite{Lottermoser, Vajk, Lonkai, Fiebig, Brown, Nandi}
more experiments supported the former one. Based on this, it was
discussed that the magnetic-field-induced ``LT1" and ``LT2" phases
are likely to have the magnetic structures of $P6_3c'm'$ and
$P6_3cm$, respectively.\cite{Lorenz} However, it has not yet been
clarified how these two magnetic transitions are driven.

Note that the $c$-axis field can hardly affect the magnetic
structure of Mn$^{3+}$ moments directly, because they are strongly
confined in the $ab$ plane. It is the significant coupling between
Mn$^{3+}$ and Ho$^{3+}$ moments that may lead to the change of
Mn$^{3+}$ sublattice when the Ho$^{3+}$ moments change their
directions or are polarized.\cite{Talbayev} This possibility was
firstly confirmed in a previous work, in which the polarization of
Ho$^{3+}$ moments formed by an external electric field causes a
90$^\circ$ rotation of Mn$^{3+}$ sublattice.\cite{Lottermoser} It
is known that the Ho$^{3+}$ ions locate on two different positions
of ``2$a$" and ``4$b$" (see Fig. 4) in the crystal lattice, but
the magnetism of Ho$^{3+}$ ions is not very
clear.\cite{Lottermoser, Brown, Hur, Nandi} Some experiments
suggested that the Ho$^{3+}$ moments on 4$b$ sites are
antiferromagnetically ordered at low temperatures and those on
2$a$ sites are disordered.\cite{Lottermoser, Hur} Apparently, the
magnetic field along the $c$ axis can induce only one sharp
polarization transition of the 4$b$ sublattice in this case.
Therefore, the two ``dip"-like transitions observed in $\kappa(H)$
curves undoubtedly demonstrate that the zero-field Ho$^{3+}$
magnetic structure must be different from that suggested
configuration.\cite{Lottermoser, Hur} The present data point to a
natural possibility as shown in Fig. 4(b), in which the Ho$^{3+}$
moments on 2$a$ and 4$b$ sublattices are all antiferromagnetically
ordered with opposite spin directions between two sublattices.
This magnetic structure was previously proposed for an
intermediate-temperature phase.\cite{Brown} Correspondingly, the
two-step sharp transitions of Ho$^{3+}$ sublattice can cause the
Mn$^{3+}$ sublattice to perform 90$^\circ$ rotations twice, with
the space group changing from $P6_3cm$ to $P6_3c'm'$ and then
probably back to $P6_3cm$ (see Fig. 4), which is consistent with
the suggestion in some earlier work.\cite{Lorenz}

\begin{figure}
\includegraphics[clip,width=8.5cm]{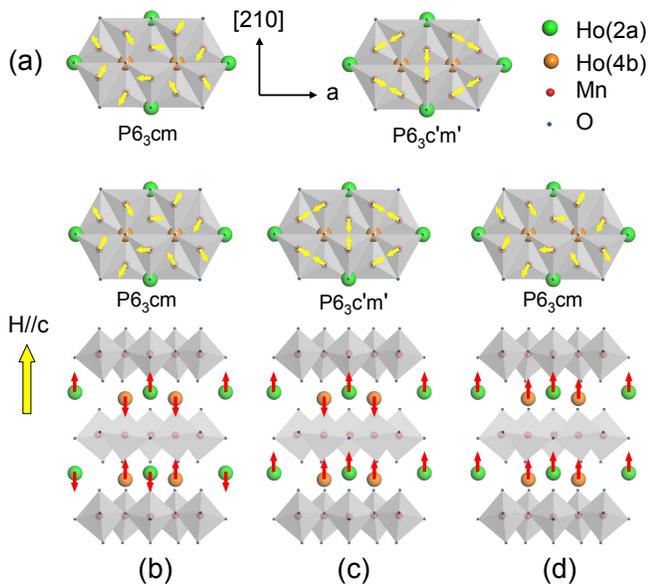}
\caption{(color online) Magnetic structure of HoMnO$_3$ at low
temperatures and those in applied magnetic field. Panel (a) shows
the zero-field ($P6_3cm$) and high-field ($P6_3c'm'$) magnetic
structures of Mn$^{3+}$ moments for $H \parallel ab$. Panels
(b--d) show the proposed two transitions of magnetic structure
when the magnetic field ($H \parallel c$) is increasing. The two
sublattices of AF Ho$^{3+}$ moments are polarized separately at
two critical fields. Correspondingly, the Mn$^{3+}$ moments make
90$^\circ$ rotations twice.}
\end{figure}

Since the low-$T$ phase diagram for $H \parallel ab$ has not been
explored,\cite{Vasic, Lemyre} the above direct comparison of heat
transport results with others is not available. In this regard,
the present heat transport data provide unprecedented information
on the low-$T$ phase diagram and phase transitions for $H
\parallel ab$. It is clear that the thermal conductivity data
demonstrate a single spin-flop transition for $H \parallel ab$,
which can only occur in the Mn$^{3+}$ sublattice considering the
Ising-like anisotropy of Ho$^{3+}$ moments. Note that in Figs.
3(a) and 3(c), we show the data for magnetic field along two
different in-plane directions, that is, [100] (along the $a$ axis)
and [210], as indicated in Fig. 4. It can be seen that there is a
clear anisotropy of the transition field between these two
directions. Similar phenomenon was observed in some other
materials, for example, the parent compounds of electron-doped
high-$T_c$ cuprate R$_2$CuO$_4$ (R = Nd, Pr),\cite{Skanthakumar,
Plakhty} in which the critical fields for their non-collinear
Cu$^{2+}$ spin structure changing to a collinear one are dependent
on the direction of magnetic field. It is not very difficult to
obtain the theoretical explanation on such anisotropy of the
spin-flop transition in the square spin lattice.\cite{Petitgrand}
However, as far as the triangular spin lattice is concerned, it is
still a remained question about how the spins rotate upon
increasing field and how such rotation depends on the direction of
the magnetic field.\cite{Chubukov} Similar to the spin
re-orientations for $H \parallel c$, it is likely that the
in-plane field may drive the zero-field magnetic structure of
$P6_3cm$ symmetry undergoing a 90$^\circ$ rotation and changing to
the $P6_3c'm'$ one at the critical field. By now, however, it is
not possible to make a straightforward judgment on which direction
is easier for the external field to drive the spin re-orientation
of the $P6_3cm$ spin structure. Nevertheless, the thermal
conductivity data indeed detect such anisotropy of transition
field in an explicit way. Finally, the proposed magnetic
structures and their transitions are summarized in Fig. 4.

\section{Summary}

The low-$T$ heat transport of HoMnO$_3$ single crystals is found
to be strongly dependent on the magnetic field, demonstrating an
exceptionally strong spin-phonon coupling. Furthermore, the
``dip"-like transitions in low-$T$ $\kappa(H)$ isotherms are found
to be useful for clarifying the magnetic structures of the
field-induced phases. The $\kappa(H)$ data for $H \parallel c$
suggest that Ho$^{3+}$ moments form two AF sublattices and undergo
two-step polarizations upon increasing field. Correspondingly, the
Mn$^{3+}$ sublattice undergoes two spin re-orientations because of
the strong interaction between Ho$^{3+}$ and Mn$^{3+}$ moments.
The $\kappa(H)$ data for $H \parallel ab$ indicate a new finding
of spin-flop transition of Mn$^{3+}$ sublattice, whose transition
field shows an in-plane anisotropy.

\section*{Acknowledgements}

We thank A. N. Lavrov and H. D. Zhou for helpful discussions. This
work was supported by the Chinese Academy of Sciences, the
National Natural Science Foundation of China, the National Basic
Research Program of China (Grant Nos. 2009CB929502 and
2006CB922005), and the Research Fund for the Doctoral Program of
Higher Education of China (Grant No. 20070358076).


\begin{thebibliography}{}

\bibitem{Lake}
B. Lake, K. Lefmann, N. B. Christensen, G. Aeppli, D. F. McMorrow,
H. M. Ronnow, P. Vorderwisch, P. Smeibidl, N. Mankorntong, T.
Sasawa, M. Nohara, and H. Takagi, Nat. Mater. {\bf 4}, 658 (2005).

\bibitem{Kang}
H. J. Kang, P. Dai, J. W. Lynn, M. Matsuura, J. R. Thompson, S.-C.
Zhang, D. N. Argyriouk, Y. Onose, and Y. Tokura, Nature (London)
{\bf 423}, 522 (2003).

\bibitem{Chang}
J. Chang, N. B. Christensen, Ch. Niedermayer, K. Lefmann, H. M.
R$\oslash$nnow, D. F. McMorrow, A. Schneidewind, P. Link, A.
Hiess, M. Boehm, R. Mottl1, S. Pailh$\acute{e}$s, N. Momono, M.
Oda, M. Ido, and J. Mesot, Phys. Rev. Lett. {\bf 102}, 177006
(2009).

\bibitem{Custers}
J. Custers, P. Gegenwart, H. Wilhelm, K. Neumaier, Y. Tokiwa, O.
Trovarelli, C. Geibel, F. Steglich, C. P$\acute{e}$pin, and P.
Coleman, Nature (London) {\bf 424}, 524 (2003).

\bibitem{Kimura}
T. Kimura, T. Goto, H. Shintani, K. Ishizaka, T. Arima, and Y.
Tokura, Nature (London) {\bf 426}, 55 (2003).

\bibitem{Aken}
B. B. Van Aken, T. T. Palstra, A. Filippetti, and N. A. Spaldin,
Nat. Mater. {\bf 3}, 164 (2004).

\bibitem{Ueland}
B. G. Ueland, J.W. Lynn, M. Laver, Y. J. Choi, and S.-W. Cheong,
Phys. Rev. Lett. {\bf 104}, 147204 (2010).

\bibitem{Lottermoser}
T. Lottermoser, T. Lonkai, U. Amann, D. Hohlwein, J. Ihringer, and
M. Fiebig, Nature (London) {\bf 430}, 541 (2004).

\bibitem{Vajk}
O. P. Vajk, M. Kenzelmann, J. W. Lynn, S. B. Kim, and S.-W.
Cheong, Phys. Rev. Lett. {\bf 94}, 087601 (2005).

\bibitem{Lorenz}
B. Lorenz, F. Yen, M. M. Gospodinov, and C. W. Chu, Phys. Rev. B
{\bf 71}, 014438 (2005).

\bibitem{Yen}
F. Yen, C. R. dela Cruz, B. Lorenz, Y. Y. Sun, Y. Q. Wang, M. M.
Gospodinov, and C. W. Chu, Phys. Rev. B {\bf 71}, 180407(R)
(2005).

\bibitem{Lemyre}
J. C. Lemyre, M. Poirier, L. Pinsard-Gaudart, and A. Revcolevschi,
Phys. Rev. B {\bf 79}, 094423 (2009).

\bibitem{Lonkai}
Th. Lonkai, D. Hohlwein, J. Ihringer, and W. Prandl, Appl. Phys.
A. {\bf 74}, S843 (2002).

\bibitem{Fiebig}
M. Fiebig, C. Degenhardt, and R. V. Pisarev, J. Appl. Phys. {\bf
91}, 8867 (2002).

\bibitem{Brown}
P. J. Brown and T. Chatterji, Phys. Rev. B {\bf 77}, 104407
(2008).

\bibitem{Nandi}
S. Nandi, A. Kreyssig, L. Tan, J.W. Kim, J. Q. Yan, J. C. Lang, D.
Haskel, R. J. McQueeney, and A. I. Goldman, Phys. Rev. Lett. {\bf
100}, 217201 (2008).

\bibitem{Hur}
N. Hur, I. K. Jeong, M. F. Hundley, S. B. Kim, and S.-W. Cheong,
Phys. Rev. B {\bf 79}, 134120 (2009).

\bibitem{Hussey}
N. E. Hussey, Adv. Phys. {\bf 51}, 1685 (2002).

\bibitem{Hess}
C. Hess, Eur. Phys. J. Special Topics {\bf 151}, 73 (2007).

\bibitem{Sologubenko}
A. V. Sologubenko, T. Lorenz, H. R. Ott, and A. Friemuth, J. Low.
Temp. Phys. {\bf 147}, 387 (2007).

\bibitem{Sun_PLCO}
X. F. Sun, I. Tsukada, T. Suzuki, S. Komiya, and Y. Ando, Phys.
Rev. B {\bf 72}, 104501 (2005).

\bibitem{Sun_GBCO}
X. F. Sun, A. A. Taskin, X. Zhao, A. N. Lavrov, and Y. Ando, Phys.
Rev. B {\bf 77}, 054436 (2008).

\bibitem{Sharma}
P. A. Sharma, J. S. Ahn, N. Hur, S. Park, S. B. Kim, S. Lee, J.-G.
Park, S. Guha, and S-W. Cheong, Phys. Rev. Lett. {\bf 93}, 177202
(2004).

\bibitem{Sun_LSCO}
X. F. Sun, S. Komiya, J. Takeya, and Y. Ando, Phys. Rev. Lett.
{\bf 90}, 117004 (2003).

\bibitem{Sun_YBCO}
X. F. Sun, K. Segawa, and Y. Ando, Phys. Rev. Lett. {\bf 93},
107001 (2004).

\bibitem{Ando_BSLCO}
Y. Ando, S. Ono, X. F. Sun, J. Takeya, F. F. Balakirev, J. B.
Betts, and G. S. Boebinger, Phys. Rev. Lett. {\bf 92}, 247004
(2004).

\bibitem{Paglione}
J. Paglione, M. A. Tanatar, D. G. Hawthorn, F. Ronning, R. W.
Hill, M. Sutherland, L. Taillefer, and C. Petrovic, Phys. Rev.
Lett. {\bf 97}, 106606 (2006).

\bibitem{Sologubenko2}
A. V. Sologubenko, K. Berggold, T. Lorenz, A. Rosch, E. Shimshoni,
M. D. Phillips, and M. M. Turnbull, Phys. Rev. Lett. {\bf 98},
107201 (2007).

\bibitem{Sun_DTN}
X. F. Sun, W. Tao, X. M. Wang, and C. Fan, Phys. Rev. Lett. {\bf
102}, 167202 (2009).

\bibitem{Berman}
R. Berman, {\it Thermal Conduction in Solids} (Oxford University
Press, Oxford, 1976).

\bibitem{Cruz}
C. dela Cruz, F. Yen, B. Lorenz, Y. Q. Wang, Y. Y. Sun, M. M.
Gospodinov, and C. W. Chu, Phys. Rev. B {\bf 71}, 060407(R)
(2005).

\bibitem{Fabreges}
X. Fabr$\grave{e}$ges, S. Petit, I. Mirebeau, S.
Pailh$\grave{e}$s, L. Pinsard, A. Forget, M. T. Fernandez-Diaz,
and F. Porcher, Phys. Rev. Lett. {\bf 103}, 067204 (2009).

\bibitem{Berggold}
K. Berggold, J. Baier, D. Meier, J. A. Mydosh, T. Lorenz, J.
Hemberger, A. Balbashov, N. Aliouane, and D. N. Argyriou, Phys.
Rev. B {\bf 76}, 094418 (2007).

\bibitem{Spin_flop}
J. A. H. M. Buys and W. J. M. de Jonge, Phys. Rev. B {\bf 25},
1322 (1982); G. S. Dixon, {\it ibid.} {\bf 21}, 2851 (1980).

\bibitem{Jin}
R. Jin, Y. Onose, Y. Tokura, D. Mandrus, P. Dai, and B. C. Sales,
Phys. Rev. Lett. {\bf 91}, 146601 (2003).

\bibitem{Talbayev}
D. Talbayev, A. D. LaForge, S. A. Trugman, N. Hur, A. J. Taylor,
R. D. Averitt, and D. N. Basov, Phys. Rev. Lett. {\bf 101}, 247601
(2008).

\bibitem{Vasic}
R. Vasic, H. D. Zhou, E. Jobiliong, C. R. Wiebe, and J. S. Brooks,
Phys. Rev. B {\bf 75}, 014436 (2007).

\bibitem{Skanthakumar}
S. Skanthakumar, J. W. Lynn, J. L. Peng and Z. Y. Li, Phys. Rev. B
{\bf 47}, 6173 (1993).

\bibitem{Plakhty}
V. P. Plakhty, S. V. Maleyev, S. V. Gavrilov, F. Bourdaro, S.
Pouget, and S. N. Barilo, Europhys. Lett. {\bf 61}, 534 (2003).

\bibitem{Petitgrand}
D. Petitgrand, S. V. Maleyev, Ph. Bourges, and A. S. Ivanov, Phys.
Rev. B {\bf 59}, 1079 (1999).

\bibitem{Chubukov}
A. V. Chubukov and D. I. Golosov, J Phys.: Condens. Matter {\bf
3}, 69 (1991).


\end{thebibliography}
\end{document}